# Solubility of fructose in water-ethanol and water-methanol mixtures by using H-bonding models

S. Tahere Alavi, Gholamreza Pazuki[*], Ahmadreza Raisi

Chemical Engineering Department, Amirkabir University of Technology (Tehran Polytechnic), Tehran, Iran

**Abstract**

The solubility of fructose in water-ethanol and water-methanol mixtures and the saturated density of each solution at 20, 30 and 40°C over a range of water mass percentage have been measured. A new modification of Wilson model was proposed to describe solubility data satisfactorily. H-bonding modified UNIQUAC and H-bonding NRTL were studied and a comparison between these three models was carried out. These three models were used to estimate new experimental data which was reported in this study. The comparison of results showed that H-bonding modified Wilson model has the lowest AAD% between these three models in the correlation of saturated density and solubility data.

## Introduction

Carbohydrate solutions have many applications in food and pharmaceutical industries. Solubility data of these solutions is needed for many process designs for example separation and purification processes. One of the most important of these processes is crystallization which is the most popular separation process after distillation. In this cases water-alcohol and alcohol-alcohol solutions are usually applied to separate sugars by precipitation. For such systems, an organic solvent acts as an anti-solvent. An anti-solvent surrounds the hydrophobic section of sugar and causes dehydration by the steric hindrance mechanism. This phenomenon reduces sugar affinity to the solvent, and consequently sugar solubility decreases and therefore an increment in crystal growth rate occurs. Fructose, glucose and sucrose are the most important sugars which are widely used in food products such as fruit juices. Among them fructose is more applicable because of its higher sweetness. Fructose that is usually with glucose in the form of HFCS (high fructose corn syrup) is purified by alcoholic crystallization.

The equilibrium data of sugar solutions are mostly limited to aqueous sugar solutions. In the case of mix solvents, the scarcity of experimental data is strongly felt. The lack of experimental data for such systems increases the need for correlation and prediction equations. Various approaches exist for the modeling of these complex systems.

Bockstanz and others (1989) measured the solubility of α-anhydrous glucose in ethanol-water mixtures at 35°C. They modeled the solubility of this system by using a Redlich-Kister series. Peres and Macedo (1997) measured the solubility of glucose in water-methanol and ethanol-methanol mixtures at 40 and 60°C. Flood and Puagsa measured the solubility, viscosity , and refractive index of fructose, glucose, ethanol and water mixture at 30°C. They also measured the densities of solutions at 25°C. Macedo and others (2000) measured the solubility of α-lactose in water-ethanol mixtures at 25, 40 and 60°C. Flood (2000) measured and modeled the solubility in the quaternary system containing fructose, glucose, ethanol and water at 30 and 40°C. He measured the solubility of sugars in a range of solvent composition from 40 weight percent ethanol to 80 weight percent ethanol. Tsavas and others (2002) measured the solubility of glucose in binary, ternary and multicomponent mixtures containing acid, esters, 2-methyl-2-butanol, dimethyl sulfoxide and water at different temperatures. Leontarakis and others (2005) measured the solubility of glucose and the equilibrium concentration of α- and β-glucose anomers in alcohols at 40 and 60°C.

Maximo and others (2010) measured the boiling point of fructose and glucose aqueous solutions. They used a modified UNIFAC model proposed by Larsen and others (1987) and showed that this model presents good accuracy for VLE prediction.

Gabas and Laguerie (1993) developed a predictive model based on the UNIFAC method. They introduced three new groups for describing sugars and calculated new interaction parameters for these groups. Catte′ and others (1994) used a modified version of the UNIQUAC model for correlating the thermodynamic properties of three binary water-sugar solutions. They used unsymmetrical convention which makes their model useless for mixed solvents. Catte′ and others (1995) also proposed a new physicochemical model to describe thermodynamic properties of binary carbohydrate-water mixtures. The chemical part of the model is related to conformational equilibria of sugars and the solvation equilibria between sugars and water. The physical part is a modified UNIFAC model, introducing three new groups. They tuned adjustable parameters of this model on seven binary water-sugar solutions. Peres and Macedo (1996) developed a modified UNIQUAC model with linear temperature-dependent parameters to describe six thermodynamic properties of three binary systems. They used the proposed model successfully to predict ternary and quaternary mixtures containing glucose, fructose, sucrose and water. Voutsas and Tassios (1997) applied a different UNIFAC-type group-contribution at the infinite-dilution region. They showed that the modified UNIFAC and entropic-FV combinatorial expressions indicate satisfactorily combinatorial effects. Peres and Macedo(1997) proposed a modified UNIFAC model. They introduced a new group "$OH_{ring}$" which considers proximity effects. This model is able to accurately predict the water activity, boiling temperature and freezing point of aqueous sucrose solution. Also, they used this model to predict vapor-liquid and solid-liquid equilibrium data for ternary and quaternary mixtures of D-glucose, D-fructose, sucrose and water using symmetric convention for all the components. Also they showed that the new model is able to predict the ternary solubility data of D-glucose, D-xylose, D-mannose and sucrose in mixtures of ethanol-water, methanol-water and methanol-ethanol. Peres and Macedo (1997) used three UNIQUAC based models- modified UNIQUAC, entropic free-volume and original Flory-Huggins- to describe the vapor-liquid and solid-liquid equilibria of aqueous solutions containing one or two sugars, as well as the solid-liquid equilibria of one sugar in mixed solvent mixtures at different temperatures. They showed that modified UNIQUAC is able to accurately correlate the ternary solubility data of D-glucose and sucrose in mixed solvent mixtures. Peres and Macedo (1999) used a modified UNIFAC model to predict the thermodynamic properties of aqueous solutions containing sugars. Also, they showed that this model is able to predict water activity of industrial sugar solutions. Spiliotis and Tassios (2000) developed a UNIFAC-type model to describe and predict aqueous and non-aqueous sugar solutions. They introduced one new main group '$CHOH_{sugar}$' in order to describe all monosaccharides which contains three subgroups. Ferreira and others (2003) used A-UNIFAC model which takes into account association effects to describe the thermodynamic properties of mixtures containing sugars, water and alcohols. They also introduced three main group to represent the sugars family. Tsavas and others (2004) modified the S-UNIFAC model which had been proposed by Spiliotis and Tassios (2000) by replacing the LLE-UNIFAC model with the modified UNIFAC model of Larsen and others (1987). They showed that this new model is applicable to predict the solubility in systems containing sugars, acids and esters. Gaida and others (2006) proposed a physical-chemical model to describe the thermodynamic properties of binary and multicomponent water-carbohydrate mixtures. They compared the results of this new model with experimental data including water activity, osmotic coefficient, activity coefficient, freezing and boiling point temperature and solubility for binary systems containing xylose, glucose, mannose, galactose, fructose, sucrose, maltose, lactose and trehalose and for food products containing sugars. The chemical part of this model dedicates to hydration equilibrium of carbohydrate with the formation of carbohydrate n-water molecules and the physical part of the model is the UNIFAC model which was modified by Larsen and others (1987). Held and others (2013) modeled thermodynamic properties and phase equilibria of aqueous sugar solutions with the PC-SAFT model. They considered 13 sugars and obtained 4 adjustable parameters from solution density and osmotic coefficient of binary sugar-water solutions at 25°C. They used these adjustable parameters to predict the sugar solubility in water and ethanol satisfactorily. They also showed that their model is able to predict the solubility and osmotic coefficient in aqueous solutions containing two solutes reasonably.

In this work, solubility of D-fructose in water-ethanol and water-methanol mixtures at 20, 30 and 40°C and also saturated density of each solution has been measured and three H-bonding models have been used to describe solid-liquid equilibria of sugar solution systems. The H-bonding modified Wilson model, the H-bonding modified UNIQUAC and the H-bonding NRTL were used to correlate saturated density and solubility of D-fructose in water-ethanol and water-methanol mixtures.

**Experimental section**

**Materials.** In all experiments double-distilled water was used. Anhydrous D-fructose was supplied by Merck Co.(Frankfurter,Germany) and no more drying was employed. Methanol and ethanol were supplied by Merck Co.(Frankfurter,Germany) with a purity of 99%.

**Method.** The solubility of D-fructose in water-ethanol and water-methanol mixtures was measured using the isothermal method which had been described by Macedo and Peres (1997). First, the mixtures of solvents with different mass fractions of water were prepared in separate cells. Then D-fructose was added to each mixtures in a small excess amount over the expected solubility. Cells were put in a shaker in which the temperature is constant. Solutions were stirred for 48 hours with a speed of 200 rpm. After stirring , solutions were put in the same temperature for about 24 hours to allow any fine suspended solid particles to settle down.

After these steps, sampling was performed by pipettes from each cell. In sampling the supernatant liquid was withdrawn and transferred into clean cells. The amount of each samples was about 2CC. The samples were weighted. In order to determine the sugar solubility of each sample, the solvents must be removed. A large amount of solvents were removed by slow drying at ambient temperature. The remaining solvents were evaporated by putting them in an oven in which the temperature was usually set at 50°C. The saturated density of each solution was measured by weighting a specific volume of solution which was withdrawn by a pipette. Each experimental point is an average of three different experiments. The standard deviation of each point is presented in tables.

**Modeling**

In this work, the equation which was used by Peres and Macedo (2000) has been applied to calculate the solubility of fructose in water-alcohol mixture. The pure sugar has been considered as the reference state for activity coefficient calculation.

$$\ln(x_i \gamma_i) = \left[ -\frac{\Delta H_f}{R} + \frac{\Delta C_p}{R}T_m + \frac{\Delta B}{2R}T_m^2 \right]\left(\frac{1}{T} - \frac{1}{T_m}\right) + \frac{\Delta C_p}{R}\ln\left(\frac{T}{T_m}\right) + \frac{\Delta B}{2R}(T - T_m) \quad (1)$$

Where

$$\Delta C_p = \Delta A + \Delta B (T - T^0) \quad (2)$$

Where $x_i$, $\gamma_i$, $\Delta H_f$, $T_m$, $T^0$ and $\Delta C_p$ are the mole fraction, the activity coefficient of the sugar, the enthalpy of fusion, the melting temperature of the sugar, an arbitrary temperature which was set equal to 298.15K and the difference between heat capacities of the pure sugar in the liquid and the solid phase, respectively. $R$ is the universal gas constant and $T$ is the mixture temperature. $\Delta A$ and $\Delta B$ are temperature independent parameters. In this work the linear equation which was used by peres and macedo (2000) has been applied for $\Delta C_p$.

For calculation of sugar solubility an appropriate model is needed to describe activity coefficient of sugar in mixture.

The local composition based model proposed by Pazuki and others (2009) was developed by the H-bonding concept. Then a comparison between three H-bonding models, the H-bonding modified Wilson, the H-bonding modified UNIQUAC and the H-bonding NRTL is carried out.

**H-bonding Modified Wilson model**

Pazuki and others (2009) proposed a modified Wilson model to calculate the activity coefficient of polymer solutions. In this work, the proposed model described by Pazuki and others is investigated.

In this model, the excess Gibbs free energy is defined as the sum of two contribution terms:

$$\frac{g^E}{RT} = \frac{g^E(combinatorial)}{RT} + \frac{g^E(residual)}{RT} \tag{3}$$

The residual term is developed as below:

$$\frac{g_{res}^E}{RT} = -C\left\{\sum_i \theta_i' \left[\ln\left(\sum_j \theta_j' H_{ji}\right)\right]\right\} \tag{4}$$

Where $C$ is the coordination factor which is set to 6 in the model. $\theta_i'$ is the modified molecular surface area fraction of component $i$ and is defined as the following relation:

$$\theta_i' = \frac{x_i q_i'}{\sum_j x_j q_j'} \tag{5}$$

$q_i'$ is the modified surface area parameter of component $i$ which is used instead of $q_i$ where hydrogen bonding exists. $H_{ij}$ is the Boltzman factor and is obtained from the interaction parameter as :

$$H_{ij} = \exp(-\frac{E_{ij}}{CRT}) \tag{6}$$

Where $E_{ij}$ is :

$$E_{ij} = h_{ij} - h_{jj} \tag{7}$$

The combinatorial term proposed by Larsen and others (1987) was used:

$$\frac{g_{comb}^E}{RT} = \sum_{i=1} x_i \ln\left(\frac{\varphi_i^{'}}{x_i}\right) \tag{8}$$

$\varphi_i^{'}$ is the modified volume fraction which was defined by Kikic and others (1980) as:

$$\varphi_i^{'} = \frac{x_i r_i^{\frac{2}{3}}}{\sum_j x_j r_j^{\frac{2}{3}}} \tag{9}$$

$r_i$ is the volume parameter of component $i$. Thus the activity coefficient of component $i$ can be expressed as:

$$\ln \gamma_i = \ln \gamma_i^{comb} + \ln \gamma_i^{res} = \ln\left(\frac{\varphi_i^{'}}{x_i}\right) + 1 - \frac{\varphi_i^{'}}{x_i} - C\left[\ln\left(\sum_j \theta_j^{'} H_{ji}\right) - 1 + \sum_j \frac{\theta_j^{'} H_{ij}}{\sum_k \theta_k^{'} H_{kj}}\right] \tag{10}$$

**H-bonding Modified UNIQUAC model**

The modified UNIQUAC model proposed by Peres and Macedo (1996) was used here and $\theta$ was replaced by $\theta^{'}$. Accordingly the activity coefficient is the sum of residual and combinatorial contributions and is defined as below:

$$\ln \gamma_i = \ln \gamma_i^{comb} + \ln \gamma_i^{res} = \ln\left(\frac{\varphi_i^{'}}{x_i}\right) + 1 - \frac{\varphi_i^{'}}{x_i} + q_i^{'}\left(1 - \ln S_i - \sum_j \frac{\tau_{ij} \theta_j^{'}}{S_j}\right) \tag{11}$$

The definition of $q_i^{'}$, $\theta_i^{'}$ and $\varphi_i^{'}$ is expressed in previous section. In this equation $S_i$ is defined as:

$$S_i = \sum_j \theta_j^{'} \tau_{ji} \tag{12}$$

Where the Boltzman factor $\tau_{ij}$ is obtained from the following equation:

$$\tau_{ij} = \exp\left(-\frac{\Delta U_{ij}}{RT}\right) \tag{13}$$

$\Delta U_{ij}$ is the interaction parameter between $i$ and $j$ :

$$\Delta U_{ij} = \frac{Z}{2}\left(U_{ij} - U_{jj}\right) \tag{14}$$

$Z$ is coordination factor and is considered 10 for the liquid phase.

## H-bonding NRTL model

In this work, $\theta$ expansion of NRTL model was used. Similar to the two other models, $\theta$ was replaced by $\theta'$ due to the existence of hydrogen bonding. Accordingly, the activity coefficient in this model can be written as:

$$\ln \gamma_i = \frac{s_i}{r_i} + \sum_j \frac{\theta'_j G_{ij}}{r_j}\left(\tau_{ij} - \frac{s_j}{r_j}\right) \tag{15}$$

Where $s_i$ and $r_i$ are expressed as below:

$$s_i = \sum_j \theta'_j \tau_{ji} G_{ji} \tag{16}$$

$$r_i = \sum_j \theta'_j G_{ji} \tag{17}$$

$G_{ij}$ and $\tau_{ij}$ are temperature dependent parameters that are obtained from interaction parameters as below:

$$\tau_{ij} = \frac{(g_{ij} - g_{jj})}{RT} \tag{18}$$

$$G_{ij} = \exp(-a_{ij}\tau_{ij}) \tag{19}$$

Linear temperature dependency for interaction parameters between sugar-water is used for the H-bonding modified Wilson and H-bonding modified UNIQUAC models. This temperature dependency was introduced by Hansen and others (1991):

$$E_{ij} = E^{(0)}_{ij} + E^{(1)}_{ij}(T - T^0) \tag{20}$$

$$\Delta U_{ij} = U^{(0)}_{ij} + U^{(1)}_{ij}(T - T^0) \tag{21}$$

$E^{(0)}_{ij}, U^{(0)}_{ij}$ and $E^{(1)}_{ij}, U^{(1)}_{ij}$ are temperature independent parameters. $T$ is the system temperature and $T^0$ is a reference temperature that was considered as 298.15K.

Structural parameters of ethanol, methanol and water are presented in Table 1. Volume parameters are calculated from the size parameters of groups which contribute to each molecules. These size parameters were taken from the UNIFAC parameter table that is presented by Gmehling and others (1982).

The modified surface area parameter of ethanol, methanol and water were calculated by Anderson and Prausnitz (1978). $q'_{fructose}$ is calculated as an adjustable parameter by Baghbanbashi(2013).

**Density modeling:**

Density of the mixture is calculated from equation 20 :

$$\ln \gamma_i = \frac{s_i}{r_i} + \sum_j \frac{\theta'_j G_{ij}}{r_j}\left(\tau_{ij} - \frac{s_j}{r_j}\right) \tag{22}$$

$x_i$, $M_i$ and $v^0_i$ are the mole fraction, the molar mass and the molar volume of component $i$, respectively and $v^E$ is the excess volume of mixture which is obtained from excess Gibbs free energy as below:

$$v^E = \left(\frac{\partial g^E}{\partial P}\right)_T \tag{23}$$

So the excess volume of mixture for the H-bonding modified Wilson model is defined as:

$$v^E = -CRT\left[\sum_i \theta'_i \ln \sum_j \theta'_j \left(\frac{\partial H_{ji}}{\partial P}\right)_T\right] \tag{24}$$

Where the derivative of $H_{ij}$ with respect to pressure at a constant temperature is:

$$\left(\frac{\partial H_{ij}}{\partial P}\right)_T = -\frac{H_{ij}}{RTC} E'_{ij} \qquad (25)$$

$E'_{ij}$ is:

$$E'_{ij} = \left(\frac{\partial E_{ij}}{\partial P}\right)_T \qquad (26)$$

The excess volume of mixture for the H-bonding UNIQUAC is calculated as below:

$$v^E = RT \sum_i \left( x_i q'_i \ln\left( \sum_j \theta'_j \left(\frac{\partial \tau_{ji}}{\partial P}\right)_T \right) \right) \qquad (27)$$

In this equation the derivative of $\tau_{ij}$ with respect to pressure at a constant temperature is defined as:

$$\left(\frac{\partial \tau_{ij}}{\partial P}\right)_T = -\frac{Z}{2} \frac{\tau_{ij}}{RT} \Delta U'_{ij} \qquad (28)$$

Where $\Delta U'_{ij}$ is:

$$\Delta U'_{ij} = \left(\frac{\partial \Delta U_{ij}}{\partial P}\right)_T \qquad (29)$$

For the H-bonding NRTL model, the excess volume of mixture is obtained from the following relation:

$$v^E = RT \sum_i \theta_i \left( \frac{\left(\frac{\partial s_i}{\partial P}\right)_T r_i - \left(\frac{\partial r_i}{\partial P}\right)_T s_i}{r_i^2} \right) \qquad (30)$$

The derivatives of $r_i$ and $s_i$ with respect to pressure at a constant temperature are:

$$\left(\frac{\partial r_i}{\partial P}\right)_T = \sum_j \theta_j \left(\frac{\partial G_{ji}}{\partial P}\right)_T \qquad (31)$$

$$\left(\frac{\partial s_i}{\partial P}\right)_T = \sum_j \theta_j \tau_{ji} \left(\frac{\partial G_{ji}}{\partial P}\right)_T + \sum_j \theta_j G_{ji} \left(\frac{\partial \tau_{ji}}{\partial P}\right)_T \qquad (32)$$

And the derivatives of temperature dependent parameters $G_{ij}$ and $\tau_{ij}$ with respect to pressure at a constant temperature are:

$$\left(\frac{\partial G_i}{\partial P}\right)_T = -\frac{a}{RT}\exp\left(-\frac{a\Delta g}{RT}\right)\left(\frac{\partial \Delta g}{\partial P}\right)_T \tag{33}$$

$$\left(\frac{\partial \tau}{\partial P}\right)_T = \frac{\left(\frac{\partial \Delta g}{\partial P}\right)_T}{RT} \tag{34}$$

In each model the derivatives of interaction parameters with respect to pressure at a constant temperature are considered as adjustable parameters.

The molecular mass and density of each component are shown in Table 1.

**Results and discussion:**

*Experimental results:*

The solubility data of fructose in the ethanol-water mixture and methanol-water mixture are presented in Table 2. Also, the standard deviation of each experimental point is shown in this table. The results reported in Table 2 indicate that the solubility of fructose increases with increasing water concentration and also with an increment in temperature. Also, the results show that the solubility of fructose in methanol-water is obviously more than the solubility of this sugar in ethanol-water mixture.

In order to validate the Solubility data which were measured in this work, a comparison between the experimental data point which were presented by Peres and Macedo (2001) and some of the solubility data of this study was carried out. The experimental solubility data which were measured by Peres and Macedo (2001) were expressed as weight percent, thus they were converted to mole percent in this work (Table 3). The comparison was carried out on the experimental data which have approximately the same conditions.

The saturated densities of these mixtures are shown in Table 4. In this table in addition to the values of saturated density, the standard deviation of each point is presented. It can be seen that the saturated density of mixture increases with an increment in temperature. This increment in the saturated density is due to the higher content of solute at higher temperature that dominates the effect of temperature increment which reduces the density of solutions.

*Thermodynamic Modeling results:*

In this study, only the interaction parameters between sugar-water, sugar-alcohol and water-alcohol, are considered as adjustable parameters. The enthalpy of fusion and the melting temperature are adopted from the literature. The fusion enthalpy and the melting temperature was measured by Raemy and Schweizer (1983) and Gabas and Laguerie (1992), respectively. $\Delta A$ and $\Delta B$ were estimated by Peres and Macedo (1997). The values of these parameters are represented in table 5. The Solubility of fructose in the mixed solvents ethanol-water and methanol-water at 20, 30 and 40°C was correlated with the studied models. Then a comparison between these three models was carried out. There are 8 adjustable parameters for each model which were fitted by the solubility data. Tables 6-8 show the interaction parameters of these three models.

In order to compare these three models, The AAD% values of each model were calculated using following equation:

$$AAD(\%) = \frac{\sum_i \left(\frac{x_j^{exp} - x_j^{calc}}{x_j^{exp}}\right)_i}{NDE} \times 100 \qquad (35)$$

Where $x^{exp}$, $x^{calc}$ and NDE are experimental mole fraction of sugar, calculated mole fraction of sugar and total number of experimental data point, respectively.

Table 9 represents the AAD% values of three models in correlation of the solubility data at each temperature. Figs.1-3 show a comparison between experimental data and correlated results of three models at different temperature for water-ethanol mixture. Figs. 4-6 show such a comparison for water-methanol mixture. These figures show that all three models correlate experimental data favorably. However, it can be seen from the figures that the H-bonding modified Wilson has better results in the correlation compared with the other two models. The reasonable results of these three models are due to the modifications which were carried out on these three local composition models in this work. These three models consider the molecular size effects. Also, $\theta$ expansion of these three models was used due to the existence of approximately large sugar molecules. Replacing $\theta'$ with $\theta$ considers the existence of hydrogen bonding which creates very powerful intermolecular forces and reduces surface area parameters. However, H-bonding modified Wilson shows better results in correlation of the solubility between other two models.

The saturated densities of fructose-ethanol-water and fructose-methanol-water mixtures which are presented in Table 4, are correlated using adjustable parameters which were calculated in previous section. Table 10 shows AAD% values of three models in correlation of the saturated density at each temperature.

It can be seen from this table that H-bonding modified Wilson has the best results in correlation of the saturated density compared with H-bonding modified UNIQUAC and H-bonding NRTL models. However it can be concluded from the values of AAD% that all three models have acceptable results in correlation of saturated density

**Conclusion:**

In the present work new experimental data on the solubility of fructose in ethanol-water and methanol-water mixtures at 20, 30 and 40°C were measured. These experimental data show that the solubility of fructose in water-methanol and water-ethanol increases with an increase of temperature. Also, an increment in water mass percentage causes higher solubility of fructose in the mixtures. On the other hand solubility of fructose in mixture of water-methanol is more than its solubility in mixture of water-ethanol at the same temperature and the same water mass percentage because hydrogen bonding between fructose and methanol is stronger than hydrogen bonding between fructose and ethanol. Thus ethanol is a better anti-solvent for separation processes. Density data show that the saturated density of mixtures increases as temperature increases because the effect of higher solubility of fructose at higher temperature dominates the inverse effect that temperature has on the density of alcohols and water.

A new modification was carried out on Wilson model to describe and predict solubility of fructose in water-methanol and water-ethanol systems. A comparison between the H-bonding modified Wilson, H-bonding modified UNIQUAC and H-bonding NRTL showed that the correlation of solubility and saturated density data with the H-bonding modified Wilson model has the minimum AAD% values between the other two models.

**Nomenclature:**

$a$ non-randomness factor

$AAD\%$ Average Absolute Deviation

$C$ coordination number

$E_{ij}$ modified Wilson interaction parameter

$g^E$ excess Gibbs energy

$g_{ij}$ NRTL interaction parameter

$h$ enthalpy

$H_{ij}$ Boltzman factor

$q'$ modified surface area parameter

$r$ volume parameter

$R$ Universal gas constant

$T$ absolute temperature (K)

$T_m$ melting temperature (K)

$T^0$ arbitrary reference temperature, set equal to 298.15 K

$U_{ij}$ UNIQUAC interaction parameter

$v^E$ excess volume of mixture

$v^0_i$ molar volume of component i

$x_i$ mole fraction of component i

$z$ coordination number

$NDE$ total number of experimental data point

**Greek letters**

$\Delta A$   temperature independent parameter ($Jmol^{-1}K^{-1}$)

$\Delta B$   temperature independent parameter ($Jmol^{-1}K^{-2}$)

$\Delta C_p$   difference between heat capacities of the pure sugar in the liquid and the solid phase

$\Delta H_f$   enthalpy of fusion

$\gamma$   activity coefficient

$\varphi'$   modified volume fraction

$\theta'$   modified molecular surface area fraction

$\tau_{ij}$   Boltzman factor

**Subscripts**

*comb*   combinatorial

*i*, *j*   specise

*res*   residual

**Superscripts**

*E*   excess

*comb*   combinatorial

*res*   residual

exp   experimental

*calc*   calculated

Table 1. Structural parameters for methanol, ethanol, water and fructose

| Component | $r_i$ | $q_i^{'}$ | $d_i^0 (g/ml)$ | $M_i (g/mol)$ |
|---|---|---|---|---|
| Fructose | 8.1529 | 1.6936 | 1.694 | 180.16 |
| Water | 0.92 | 1 | 1 | 18.02 |
| Ethanol | 2.5755 | 0.92 | 0.789 | 46.07 |
| Methanol | 1.9011 | 0.96 | 0.792 | 32.04 |

Table 2. Experimental solubility of fructose in water-ethanol and water-methanol mixtures(mole/mole)%

| SR[a] \ T(°C) | 20% | Σ | 40% | Σ | 60% | σ | 80% | σ | 90% | Σ | 100% | Σ |
|---|---|---|---|---|---|---|---|---|---|---|---|---|
| **Water-Ethanol** | | | | | | | | | | | | |
| 20°C | 7.50[b] | 0.006 | 17.80 | 0.003 | 23.30 | 0.004 | 25.50 | 0.007 | 27.44 | 0.001 | 28.19 | 0.002 |
| 30°C | 8.30 | 0.002 | 18.55 | 0.003 | 24.37 | 0.003 | 26.80 | 0.008 | 28.10 | 0.006 | 29.33 | 0.005 |
| 40°C | 15.15 | 0.006 | 25.26 | 0.007 | 28.80 | 0.005 | 31.77 | 0.004 | 32.50 | 0.003 | 33.49 | 0.005 |
| **Water-Methanol** | | | | | | | | | | | | |
| 20°C | 10.4[b] | 0.001 | 15.80 | 0.001 | 21.40 | 0.001 | 23.50 | 0.005 | 24.96 | 0.004 | 27.60 | 0.004 |
| 30°C | 15.90 | 0.004 | 24.40 | 0.002 | 27.86 | 0.003 | 28.20 | 0.001 | 29.60 | 0.007 | 30.16 | 0.005 |
| 40°C | 19.70 | 0.008 | 28.70 | 0.003 | 29.46 | 0.003 | 32.28 | 0.006 | 33.68 | 0.003 | 34.40 | 0.004 |

a. SR(solution Ratio) is mass percentage of water in the sugar free solvent b. Solubility is expressed as mole percentage of fructose in mixture

Table 3. A comparison between the literature experimental data and the experimental data point which were measured in this study. (the gray sections are the literature experimental data)

| SR(a) T(°C) | 20% | 20.099% | 40% | 39.911% | 60% | 59.905% | 80% | 79.428% |
|---|---|---|---|---|---|---|---|---|
| **Water-Ethanol** | | | | | | | | |
| **40°C** | 15.15 | 15.7 | 25.26 | 26.3 | 28.80 | 32 | 31.77 | 34 |
| **Water-Methanol** | | | | | | | | |
| | 20% | 19.93% | 40% | 39.970% | 60% | 58.732% | 80% | 80.46% |
| **40°C** | 19.7 | 20.95 | 28.7 | 29.3 | 29.46 | 32.3 | 32.28 | 34.2 |

a. SR(solution Ratio) is mass percentage of water in the sugar free solvent

Table 4. Saturated density of fructose-ethanol-water and fructose-methanol-water mixtures

| SR[a] T(°C) | 20% | σ | 40% | σ | 60% | σ | 80% | σ | 90% | σ | 100% |
|---|---|---|---|---|---|---|---|---|---|---|---|
| **Fructose-Water-Ethanol** | | | | | | | | | | | |
| 20°C | 0.801[b] | 0.007 | 0.963 | 0.013 | 1.151 | 0.016 | 1.214 | 0.004 | 1.291 | 0.003 | 1.302 |
| 30°C | 0.821 | 0.01 | 1.104 | 0.008 | 1.231 | 0.014 | 1.304 | 0.008 | 1.316 | 0.01 | 1.321 |
| 40°C | 0.902 | 0.006 | 1.115 | 0.005 | 1.267 | 0.009 | 1.313 | 0.002 | 1.336 | 0.009 | 1.368 |
| **Fructose-Water-Methanol** | | | | | | | | | | | |
| 20°C | 0.976[b] | 0.003 | 1.138 | 0.006 | 1.256 | 0.009 | 1.292 | 0.005 | 1.301 | 0.003 | 1.314 |
| 30°C | 1.008 | 0.007 | 1.153 | 0.011 | 1.269 | 0.004 | 1.317 | 0.013 | 1.320 | 0.01 | 1.328 |
| 40°C | 1.117 | 0.012 | 1.287 | 0.018 | 1.319 | 0.010 | 1.343 | 0.009 | 1.352 | 0.019 | 1.372 |

a. SR(solution Ratio) is mass percentage of water in the sugar free solvent  b. Density is reported as g/ml.

Table 5. Thermodynamic data used for the calculation of the solubility of fructose in water-alcohol mixtures

|  | D-fructose |
|---|---|
| Melting Temperature(K) | 378.15[1] |
| Fusion enthalpy(J/mol) | 32428[2] |
| $\Delta A$ (J/mol.K) | 126.1469[3] |
| $\Delta B$ (J/mol.K$^{-2}$) | 0[3] |

(1) Gabas and Laguerie(1992) (2) Raemy and Schweizer (1983) (3) Peres and Macedo (1997)

Table 6.Interaction parameter (H-bonding modified Wilson model)

| | Fructose | Ethanol | Water |
|---|---|---|---|
| Fructose | 0 | $-9.408 \times 10^3$ | $-3.882 \times 10^{3(a)}$ |
| | | | $0.632 \times 10^{3(b)}$ |
| Ethanol | $3.4201 \times 10^4$ | 0 | $8.135 \times 10^3$ |
| Water | $-0.253 \times 10^{3\ (a)}$ | $-9.415 \times 10^3$ | 0 |
| | $-0.498 \times 10^{3\ (b)}$ | | |

| | Fructose | Methanol | Water |
|---|---|---|---|
| Fructose | 0 | $-8.969 \times 10^3$ | $-1.7028 \times 10^{4\ (a)}$ |
| | | | $-0.82 \times 10^{2(b)}$ |
| Methanol | $1.9421 \times 10^4$ | 0 | $1.689 \times 10^3$ |
| Water | $4.92542 \times 10^{5\ (a)}$ | $-0.44 \times 10^2$ | 0 |
| | $7.329 \times 10^{3(b)}$ | | |

a. $E_{ij}^{(0)}$ in Kelvin  b. $E_{ij}^{(1)}$

Table 7. Interaction parameter (H-bonding modified UNIQUAC)

|  | Fructose | Ethanol | Water |
|---|---|---|---|
| Fructose | 0 | $0.732 \times 10^3$ | $-0.227 \times 10^{3(a)}$ $-50^{(b)}$ |
| Ethanol | $2.138 \times 10^3$ | 0 | 110 |
| Water | $-0.201 \times 10^{3(a)}$ $51^{(b)}$ | $0.343 \times 10^3$ | 0 |

|  | Fructose | Methanol | Water |
|---|---|---|---|
| Fructose | 0 | $-0.340 \times 10^3$ | $-1.190 \times 10^{3(a)}$ $-8^{(b)}$ |
| Methanol | $2.1945 \times 10^4$ | 0 | $-1.039 \times 10^3$ |
| Water | $8.3836 \times 10^{4(a)}$ $1.5092 \times 10^{4\ (b)}$ | $0.505 \times 10^3$ | 0 |

a. $U_{ij}^{(0)}$ in Kelvin  b. $U_{ij}^{(1)}$

Table 8.Interaction parameter (H-bonding NRTL)

|  | Fructose | Ethanol | Water |
| --- | --- | --- | --- |
| Fructose | 0 | $1.252\times10^3$ | $-6.140\times10^3$ |
| Ethanol | $1.2666\times10^4$ | 0 | $-5.587\times10^3$ |
| Water | $-0.860\times10^3$ | $8.090\times10^3$ | 0 |

|  | Fructose | Methanol | Water |
| --- | --- | --- | --- |
| Fructose | 0 | $3.907\times10^3$ | $-1.910\times10^3$ |
| Methanol | $-2.421\times10^3$ | 0 | $-4.917\times10^3$ |
| Water | $-3.769\times10^3$ | $1.1299\times10^4$ | 0 |

Table 9.The AAD% values for different models in correlation of solubility data

| System | T(°C) | AAD% H-bonding modified Wilson | AAD% H-bonding modified UNIQUAC | AAD% H-bonding NRTL |
|---|---|---|---|---|
| Fructose-Water-Ethanol | 20°C | 1.74 | 4.34 | 8.68 |
| | 30°C | 2.18 | 1.33 | 10.19 |
| | 40°C | 2.06 | 4.05 | 2.63 |
| Fructose-Water-Methanol | 20°C | 5.16 | 5.03 | 6.07 |
| | 30°C | 5.20 | 8.05 | 7.79 |
| | 40°C | 4.71 | 4.76 | 3.22 |
| **Average** | | **3.50** | **4.59** | **6.43** |

Table 10.The AAD% values for different models in correlation of saturated density.

| System | T(°C) | AAD% H-bonding modified Wilson | AAD% H-bonding modified UNIQUAC | AAD% H-bonding NRTL |
|---|---|---|---|---|
| Fructose-Water-Ethanol | 20°C | 1.30 | 1.33 | 1.06 |
| | 30°C | 1.07 | 1.71 | 0.45 |
| | 40°C | 0.56 | 0.67 | 1.15 |
| Fructose-Water-Methanol | 20°C | 0.60 | 1.58 | 1.20 |
| | 30°C | 0.46 | 0.68 | 2.32 |
| | 40°C | 0.85 | 1.33 | 0.91 |
| **Average** | | **0.81** | **1.21** | **1.18** |

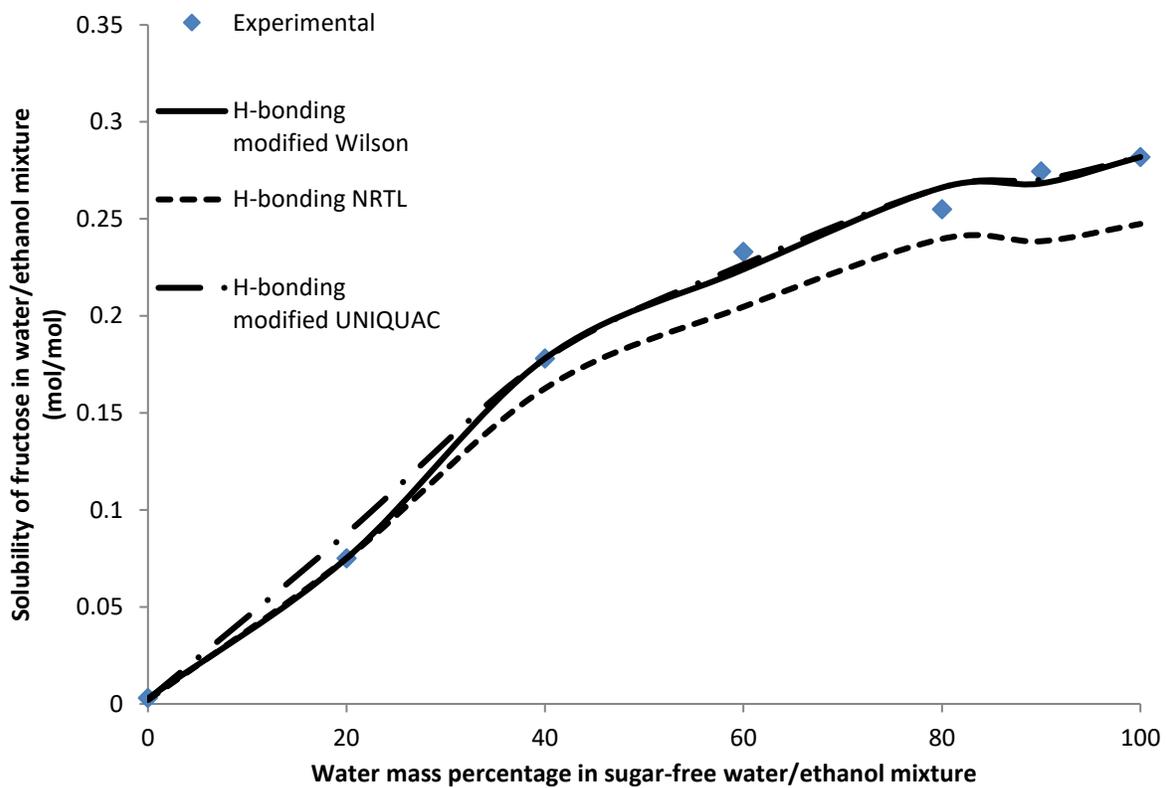

**Fig. 1**. Experimental and correlated results of solubility for fructose-water-ethanol mixture at 20°C.

Figure 1

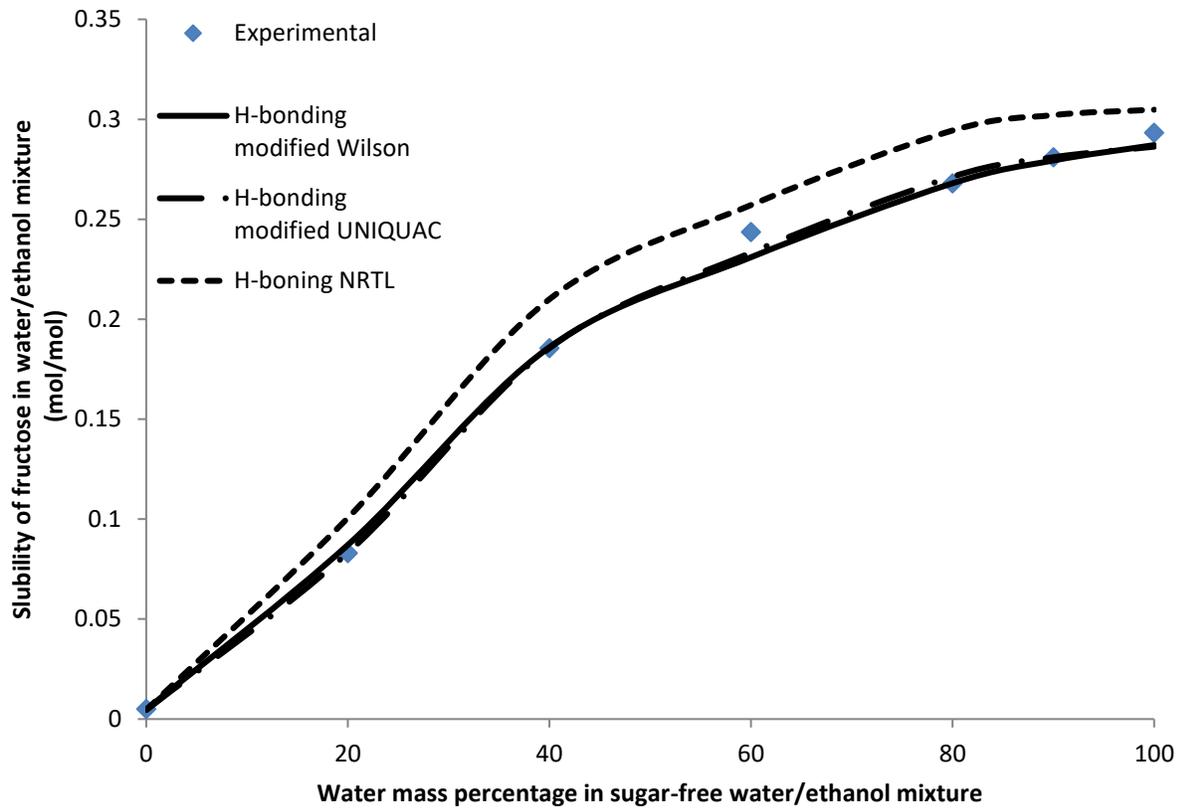

**Fig. 2**.Experimental and correlated results of solubility for fructose-water-ethanol mixture at 30°C.

Figure 2

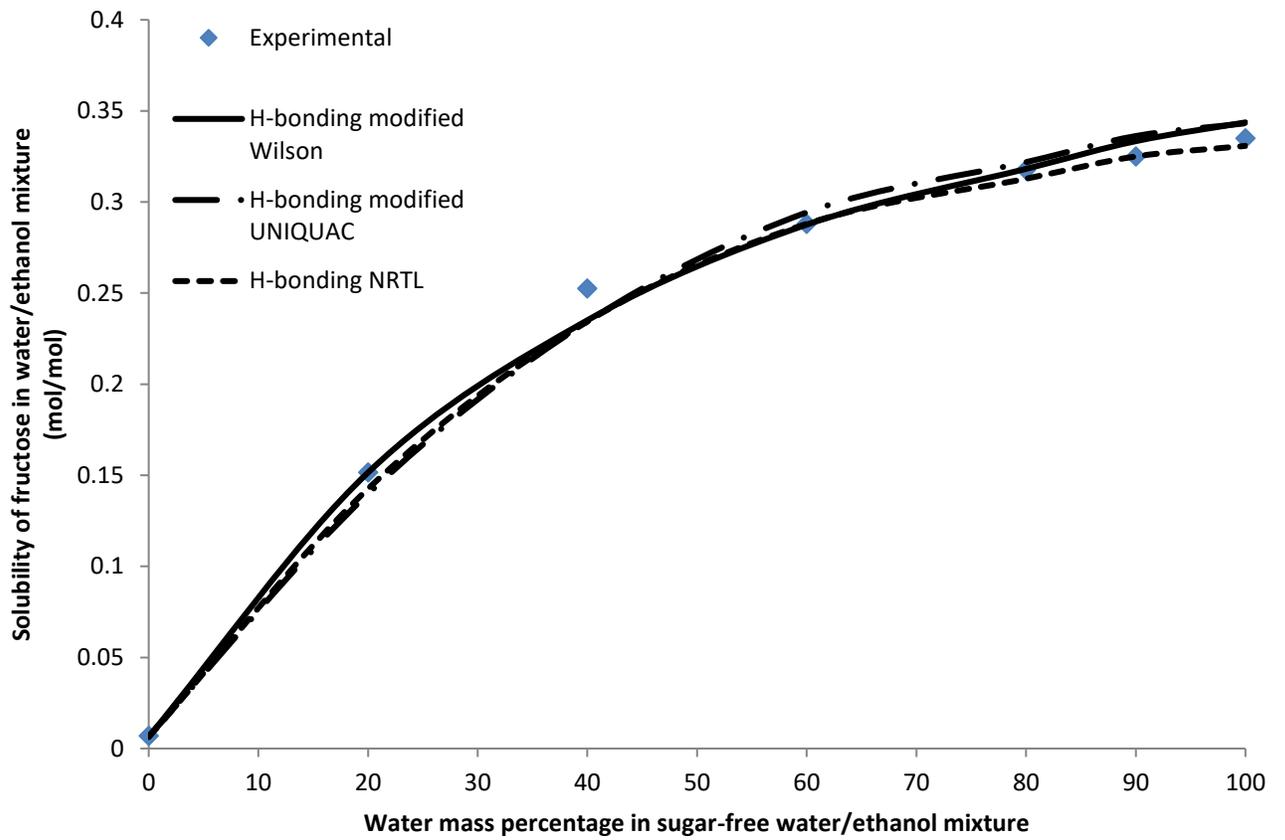

Fig. 3.Experimental and correlated results of solubility for fructose-water-ethanol mixture at 40°C.

Figure 3

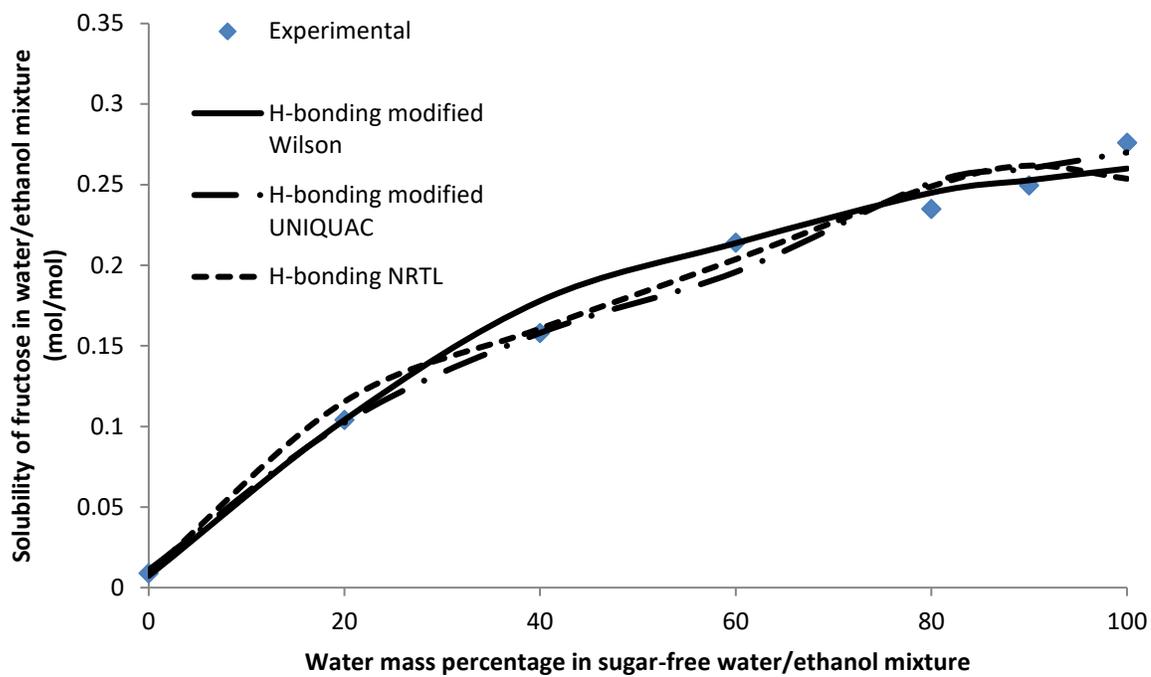

Fig. 4.Experimental and correlated results of solubility for fructose-water-methanol mixture at 20°C.

Figure 4

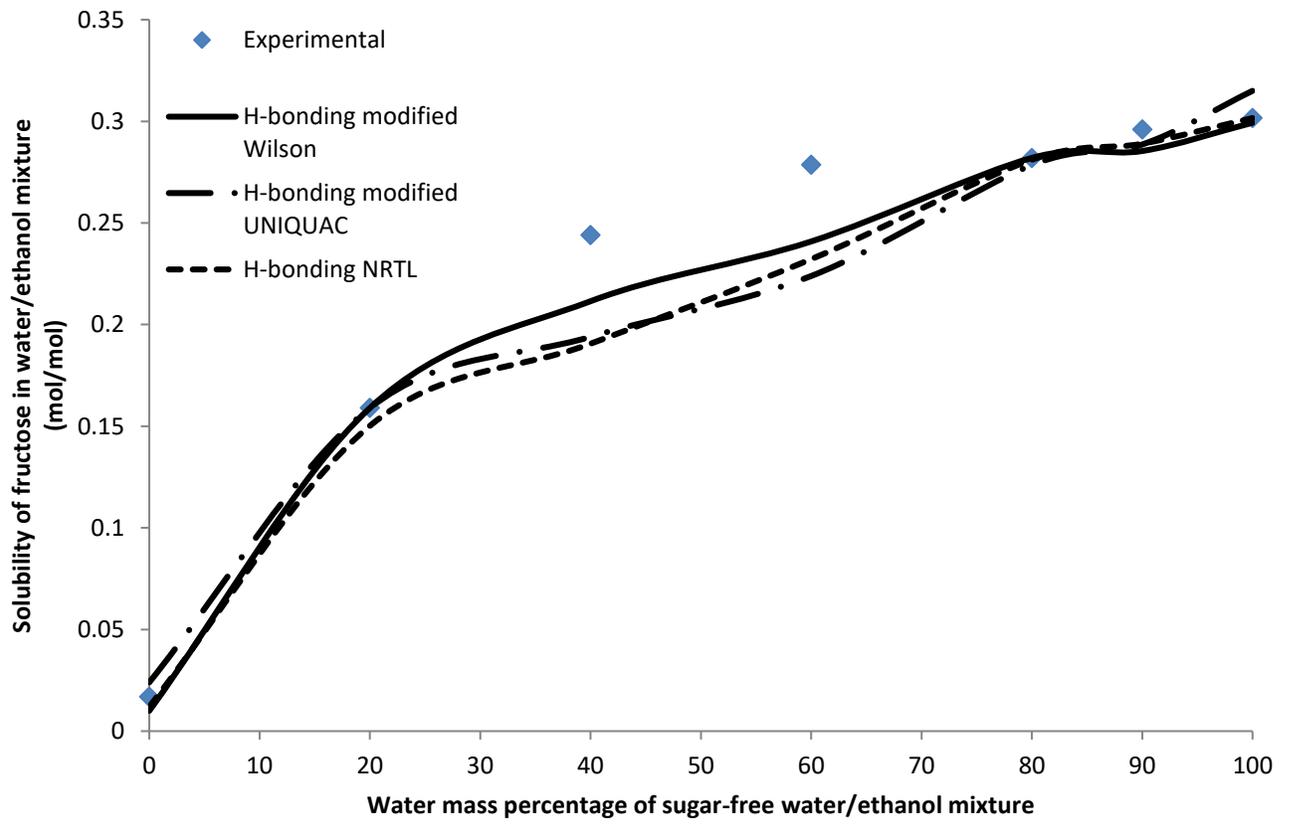

Fig. 5. Experimental and correlated results of solubility for fructose-water-methanol mixture at 30°C.

Figure 5

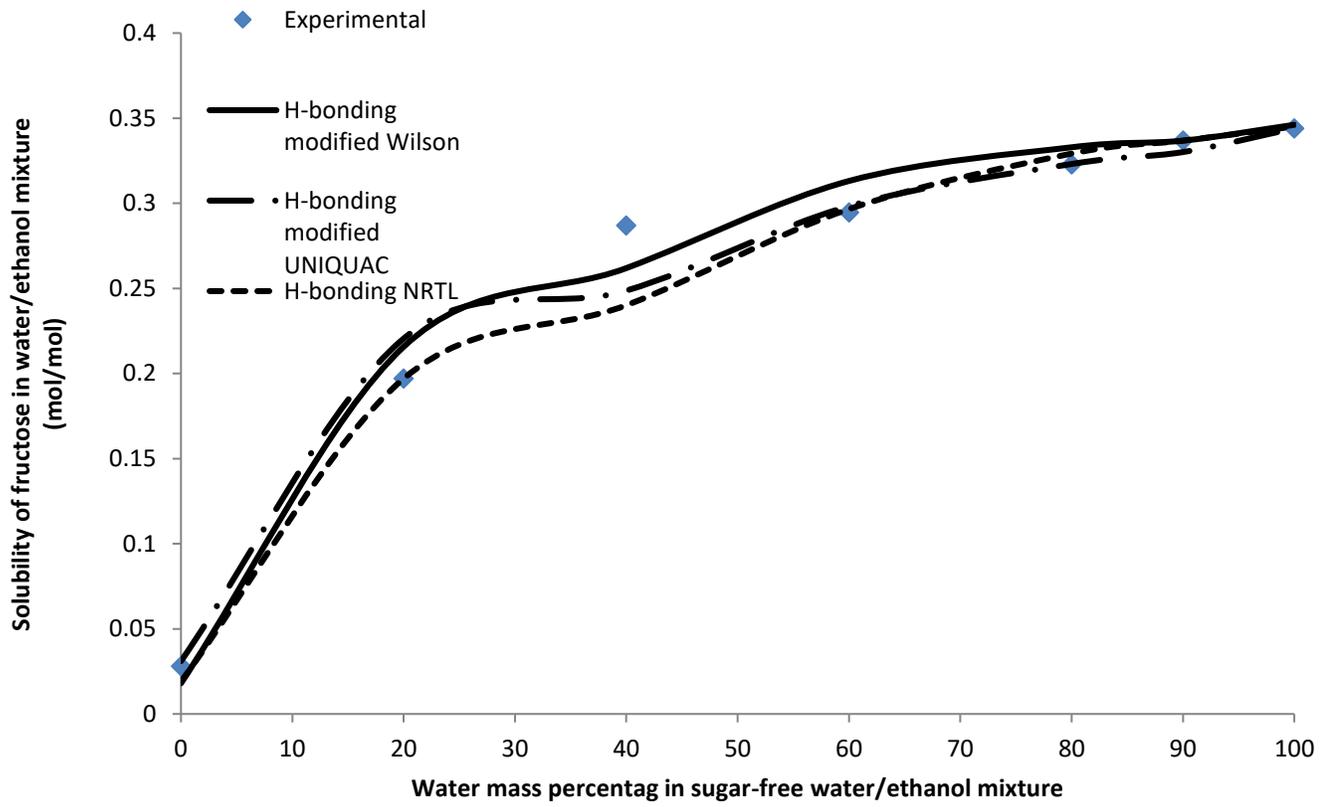

Fig. 6. Experimental and correlated results of solubility for fructose-water-methanol mixture at 40°C.

Figure 6